\documentclass[11pt,fleqn]{article}

\input epsf
\newcommand{\beq}{\begin{equation}}
\newcommand{\eeq}{\end{equation}}
\newcommand{\ba}{\begin{eqnarray}}
\newcommand{\ea}{\end{eqnarray}}
\renewcommand{\a}{\alpha}

\newcommand{\m}{\mu}

\newcommand{\rmd}{{\rm d}}
\newcommand{\llr}{\langle\lambda\rangle}
\newcommand{\lthetar}{\langle\theta\rangle}
\newcommand{\lNr}{\langle\hat{N}\rangle}
\begin{document}

\title{\normalsize \hfill DO-TH 98/21 \\
\normalsize \hfill CBPF-NF-080/98\\ 
\LARGE 
\bf On the Free-Energy of Three-Dimensional CFTs and Polylogarithms}

\author{{\bf{Anastasios C. Petkou}} 
         \thanks{email: petkou@catbert.physik.uni-dortmund.de} 
	\thanks{Address after 1st December 1998: Department of
         Theoretical Physics,  Aristotle University of
         Thessaloniki, Thessaloniki 54006, Greece.}\\
         \small Institut f\" ur Physik, Universit\"at Dortmund,
         D-44221, Germany \\ 
        and \\ 
         {\bf Marcello B. Silva Neto} 
         \thanks{e-mail: sneto@lafex.cbpf.br}  \\
        \small Centro Brasileiro de Pesquisas F\'{\i}sicas, Rua Dr. Xavier
         Sigaud 150 - Urca \\ \small CEP:22290-180, Rio de Janeiro, Brazil} 

\maketitle

\begin{abstract}
\noindent
We study the $O(N)$ vector model 
and the $U(N)$ Gross-Neveu model with fixed total fermion number, in
three dimensions. Using        
non-trivial polylogarithmic identities, we calculate the large-$N$ renormalized
free-energy density of these models, at their  conformal points in a
``slab'' geometry with one finite dimension of length $L$.  We
comment on the possible implications of our results. 

\end{abstract} 

\newpage

\section{Introduction}

Conformal field theories (CFTs) in dimensions $d>2$ have recently
attracted much interest \cite{Maldacena}. Generic results
based solely on conformal invariance are not very restrictive in $d>2$, (see 
however \cite{Todorov,Fradkin,Tassos1}), and most of the information is 
extracted by studying explicit models. Although many CFTs have been
recently 
discovered in $d=4$ \cite{Seiberg}, a large amount of work has also been 
devoted to the study of CFT models in $2<d<4$, such as the $O(N)$
vector model \cite{Vasiliev,Ruehl,Tassos2}  
and the Gross-Neveu model \cite{Gracey} at their conformal points for
large-$N$.

In this work, we study the $O(N)$
vector model and  
the $U(N)$ Gross-Neveu model with fixed total fermion number, in $d=3$.  
Using non-trivial polylogarithmic
identities, we calculate the free-energy density of these models to
leading order in the $1/N$ expansion, at their
conformal points in a ``slab'' geometry with one finite dimension of
length $L$.  The free-energy density of the three-dimensional $O(N)$
vector model in a ``slab'' geometry was first calculated in
\cite{Sachdev1,Sachdev2}. Recently, the free-energy density of
(super)conformal field theories in $d=4$ was also calculated in terms
of polylogarithms \cite{Gubser}. 

Based on earlier ideas \cite{Castro-Neto}, it has been recently
argued in \cite{Appelquist} that the free-energy density of four dimensional
quantum field theories (QFTs) encodes non-perturbative information for
the massless degrees of freedom coupled to a fixed point. Therefore,
it may be interpreted as a measure of their expected reduction from the
UV to the IR \cite{Zamolodchikov}. Our leading-$N$ analytic results
are consistent 
with an extension of such an interpretation for the free-energy density
to three-dimensional QFTs.

\section{The Conformally Invariant $O(N)$ Vector Model in $d=3$} 
We begin by reviewing the results of \cite{Sachdev2}. Consider the partition 
function of the $O(N)$ vector model in $d=3$, obtained after integrating out
the fundamental scalar fields $\phi^{\a}(x)$, $\a =1,2,..,N$,  
\ba Z_{B} &=&\int ({\cal {D}}\sigma )\,\exp
\left[ -N\,S_{eff}(\sigma ,g)\right]\,, \label{eq1}\\
S_{eff}(\sigma ,g)&=&\frac{1}{2}\mbox{Tr}[\ln (-\partial ^{2}+\sigma
)]-\frac{1}{2g} \int {\rm d}^{3}x\,\sigma(x)\label{eq2}\,, 
\ea 
where $\sigma(x)$ is an auxiliary scalar field and $g$ is the coupling.
Setting $\sigma (x)=M^{2} + ({\rm i}/\sqrt{N})\,\sigma_{1} (x)$,
(\ref{eq1}) can be 
calculated in a renormalisable $1/N$ expansion \cite{Rosenstein1},
provided the gap equation 
\begin{equation}
\frac{1}{g}=\int \frac{{\rm d}^{3}p}{(2\pi
)^{3}}\frac{1}{p^{2}+M^{2}}\,, \label{eq3}
\end{equation}
is satisfied. To any fixed order in $1/N$, a non-trivial CFT is obtained by 
tuning the coupling to the critical value 
$1/g \equiv 1/g_{\ast }=(2\pi)^{-3}\int{\rm d}^{3}p/p^{2}$ \cite{Rosenstein1}.
Then, the renormalised mass (or inverse correlation length) $M(= 1/ \xi)=0$.

When the model is placed in a ``slab'' geometry with one finite dimension
of length $L$ and periodic boundary conditions, the gap equation reads
\beq  
\frac{1}{g}
 = \frac{1}{L}\,\!\!\sum\limits_{n=-\infty }^{\infty }\int \frac{
{\rm d}^{2}{\bf{p}}}{(2\pi
)^{2}}\frac{1}{{\bf{p}}^{2}+\omega_{n}^{2}+M^{2}_{L}}\,, \hspace{2cm}
{\bf{p}}=(p_{1},p_{2})\,, \label{eq4} 
\eeq
with the momentum along the finite dimension taking the values
$\omega_{n}=2\pi n/L$, $n=0,\pm 1,\pm 2,..$. We may then study the
existence of  
a finite-temperature phase transition in the dimensionally continued
version of  
the model, (i.e. in $d-1$ infinite dimensions)
\cite{Rosenstein1,Kogut,Ruehl2}.  
Since UV renormalisation is insensitive to putting the system in a
finite geometry, the coupling  
constant on the l.h.s. of (\ref{eq4}) can be set to its renormalised value in 
the bulk which explicitly depends on a  mass scale $M_{0}$, such that 
the system is in its 
$O(N)$-ordered  phase for zero ``temperature'' $T\sim 1/L$. Then, we
can obtain 
an equation which gives the dependence of $M_{L}$ on $M_{0}$ and $T$. The
finite-temperature phase transition corresponding to $O(N)$ symmetry
restoration,  occurs when $M_{L}=0$ for some critical 
temperature $T_{*}$ which is related to $M_{0}$ through the
above equation. It  
can be shown that the finite-temperature phase transition cannot
take place for
$2<d\leq 3$ in accordance to the Mermin-Wagner-Coleman theorem, but can only 
occur for $3<d<4$. The critical temperature $T_{*}$  obtained this way
agrees \cite{Tassos3} with the well-known results, i.e. see
\cite{Svaiter}. The  
theory at $T_{*}$ is a $d-1$-dimensional CFT, however the OPE structure of its 
correlation functions is less clear \cite{Tassos3}.

On the other hand, when the coupling is fixed to its bulk critical value 
$1/g_{*}$, (\ref{eq4}) has the solution
\beq
M_{L}\equiv
M_{*}=\frac{2}{L}\ln\left(\frac{1+\sqrt{5}}{2}\right)\,,\label{eq5} 
\eeq
which corresponds to the physical situation of finite-size scaling 
\cite{Cardy1} of the correlation length. Then, the subtracted
\footnote{Since the UV 
singularities are the same in the bulk and in the finite geometry,
subtraction of the bulk free-energy density ensures a UV-finite result.}
free-energy density for this configuration reads
\ba
\frac{f_{\infty}-f_{L}}{N} & = &
\frac{1}{2}\int\frac{\rmd^{3}p}{(2\pi)^{3}} \ln
p^{2}-\frac{1}{2L}\sum_{n=-\infty}^{\infty}
\int\frac{\rmd^{2}{\bf{p}}}{(2\pi)^{2}}\ln({\bf{p}}^{2}+
\omega_{n}^{2}+M_{*}^{2}) 
+\frac{1}{2}\int\frac{\rmd^{3}p}{(2\pi)^{3}}\frac{M_{*}^{2}}{p^{2}}
\nonumber \\
 & = & \frac{M_{*}^{3}}{12\pi}-\frac{1}{2\pi L^{3}}\left[\ln({\rm
e}^{-LM_{*}}) Li_{2}({\rm e}^{-LM_{*}})-Li_{3}({\rm
e}^{-LM_{*}})\right] \nonumber \\
& = & \frac{4}{5}\frac{\zeta(3)}{2\pi L^{3}}\,,\label{eq6}
\ea  
where $Li_{n}(x)$ are the usual polylogarithms \cite{Lewin}. The
third line in (\ref{eq6}) follows from the second, by virtue
of non-trivial polylogarithmic identities \cite{Lewin}, and fits into the 
general formula \cite{Cardy2} 
\beq
f_{\infty}-f_{L}=\tilde{c}\,\frac{2\zeta(d)}{S_{d}L^{d}}\,,\label{eq7}
\eeq
for the finite-size scaling of the free-energy density in conformal
field theories, with $\tilde{c}/N=4/5$. It is quite remarkable that
the value of $\tilde{c}/N$ obtained from (\ref{eq6}), turns out to be a
rational number. This is reminiscent of central charge \footnote{Recall
that, in two dimensions $\tilde{c}$ coincides with the central charge
\cite{Affleck}.}  
calculations in 
two-dimensional CFTs, (see \cite{Jose} for a recent reference). 
There is strong evidence
\cite{Tassos4} that 
correlation functions at the above finite-size critical point can by
described by operator 
product expansions of the bulk $O(N)$ vector model,
in accordance with earlier ideas \cite{Cardy2}. 

The result (\ref{eq6}) is consistent with the interpretation of
$\tilde{c}$ as a measure of the massless degrees of freedom coupled to a
critical point and their expected reduction from the UV to the
IR. To see this, we recall that the large-$N$ critical point of the $O(N)$
vector model is identical with the large-$N$ IR critical point of the $O(N)$
invariant $\phi^{4}$ theory \cite{Zinn}. The latter
theory has as UV critical limit the free theory of $N$ massless
bosons, for which it is well-known that $\tilde{c}/N=1$ in
$d=3$. Therefore, according to the interpretation given in
\cite{Appelquist}, we expect that 
\beq
\tilde{c}_{UV}=\tilde{c}(N\mbox{ massless free bosons})>
\tilde{c}(O(N) \mbox{vector 
model}) = \tilde{c}_{IR} \,,\label{new1}
\eeq
which is satisfied since $1>4/5$. Note that (\ref{new1}) is a
non-perturbative result from the point of view of the renormalisation
group (RG) flow, since the large-$N$ IR critical point of the
$\phi^{4}$ theory is non-perturbative in the coupling.  

\section{The $U(N)$ Gross-Neveu Model in $d=3$} 

Following the considerations above, it is possible to
study the free-energy density of other 
CFT models. Consider, for example, the 
$U(N)$ invariant Gross-Neveu model in $d=3$ whose partition function, after
integrating out the fundamental Dirac fermionic fields $\psi^{\a}(x)$,
$\bar{\psi}^{\a}(x)$, $\a=1,2,..,N$, reads
\ba 
Z_{F} &=&\int ({\cal {D}}\lambda )\,\exp
\left[ -N\,I_{eff}(\lambda ,G)\right]\,, \label{eq8}\\
I_{eff}(\lambda ,G)&=&\frac{1}{2G} \int {\rm d}^{3}x\,\lambda^{2}
(x)-\mbox{Tr}[\ln (\slash\!\!\!\partial+\lambda 
)]  \label{eq9}\,, 
\ea
where $\lambda(x)$ is an auxiliary scalar field and $G$ is the
coupling. We use the notation $\slash\!\!\!\partial=\gamma_{\m}\partial_{\m}$ 
and the following two-dimensional Hermitian representation for the Euclidean
gamma matrices  in $d=3$ \cite{Rosenstein2,Zinn}
\beq
\gamma_{1}=\sigma^{1}\,,\hspace{1.2cm}\gamma_{2}=\sigma^{2}\,,\hspace{1.2cm}
\gamma_{3}\equiv\gamma_{0}=\sigma^{3}\,, \label{eq10}
\eeq
where $\sigma^{i}$, $i=1,2,3$ are the usual Pauli matrices. This model 
describes fermion mass generation through the breaking of space parity
\cite{Zinn}. The partition function (\ref{eq8}) can be 
evaluated in a renormalisable $1/N$ expansion \cite{Rosenstein2} when one 
sets $\lambda(x)=m+(1/\sqrt{N})\,\lambda_{1}(x)$, provided the following gap
equation is satisfied
\beq
\frac{1}{G}=2\int\frac{\rmd^{3}p}{(2\pi)^{3}}\frac{1}{p^{2}+m^{2}}
\,.\label{eq11}
\eeq
At the critical coupling $1/G=1/G_{*}=2(2\pi)^{-d}\int{\rmd}^{d}p/p^{2}$, the 
theory is conformally invariant and $m=0$. 

When the system is placed in a ``slab'' geometry with one finite dimension of 
length $L$, the fermions acquire antiperiodic boundary 
conditions and the gap equation reads
\beq  
\frac{1}{G}
 = \frac{2}{L}\,\!\!\sum\limits_{n=-\infty }^{\infty }\int \frac{
{\rm d}^{2}{\bf{p}}}{(2\pi
)^{2}}\frac{1}{{\bf{p}}^{2}+\omega_{n}^{2}+m^{2}_{L}}\,, \label{eq12} 
\eeq  
with $\omega_{n}=(2n+1)\pi/L$, $n=0,\pm 1,\pm 2,..$. Again, we may study
the finite-temperature phase transition in the dimensionally continued
version of the model, (i.e. in $d-1$ infinite dimensions) 
\cite{Rosenstein2,Kogut}, by setting $1/G$ to its bulk renormalised
value which explicitly depends on the mass $m_{0}$ of the fundamental
fermionic fields. This means that the system is in its ``broken
phase'' for  zero 
``temperature'' $T\sim 1/L$. Then, we can obtain an equation which gives the 
dependence of $m_{L}$ on $m_{0}$ and $T$. The second order 
finite-temperature phase transition corresponding to space parity
restoration,  occurs when $m_{L}=0$ for some critical 
temperature $T_{*}$ which is related to 
$m_{0}$ through the above equation.  
The finite-temperature phase transition is now possible for all $2<d<4$ 
\cite{Rosenstein2,Kogut}, due to the absence of zero modes for 
fermions and antiperiodic boundary conditions.

On the other hand, when the coupling stays at its bulk critical value
$1/G_{*}$, (\ref{eq12}) is 
satisfied for
\beq 
m_{L}\equiv m_{*}=0. 
\label{fermion-zero-mode}
\eeq
This essentially means that, to leading order in $1/N$, the free-energy density
of the system is   
given by the free-field theory result.\footnote{There exist solutions
of (\ref{eq12}) with imaginary $m_{L}$ which will be discussed
elsewhere.} Indeed we easily find
\beq
\frac{f_{\infty}-f_{L}}{N}=
\frac{3}{2}\frac{\zeta(3)}{2\pi\,L^{3}}\,,\label{eq13} 
\eeq
which implies that $\tilde{c}/N=3/2$ in agreement with the results of
\cite{Vitale}. 

Although the result (\ref{eq13}) may seem trivial, it encodes
important information. This is again seen if we recall that the
large-$N$ critical point of the $U(N)$ Gross-Neveu model is identical
with the large-$N$ IR critical point of the Gross-Neveu-Yukawa model
\cite{Zinn}. The latter model has as UV critical 
point the free theory of $N$ massless Dirac fermions plus one massless
boson for which $\tilde{c}/N=3/2+1/N$ in $d=3$. Then, according to
\cite{Appelquist} we expect
\beq
\tilde{c}_{UV}=\tilde{c}(N\mbox{ massless Dirac fermions + 1 massless
boson}) > 
\tilde{c}(\mbox{Gross-Neveu})=\tilde{c}_{IR}\,,\label{new2}
\eeq
which is consistent with (\ref{eq13}) for large and  finite $N$. Again, the
non-perturbative nature of (\ref{new2}) is a consequence of the
non-perturbative nature of the large-$N$ IR critical point of the
Gross-Neveu-Yukawa model.  

\section{The $U(N)$ Gross-Neveu Model in $d=3$ with fixed total fermion number}
 
The Gross-Neveu model can also be studied \cite{Weiss,Marcello} for
fixed total  
fermion number $B$. To this effect, we introduce a delta-function constraint 
$\delta(\hat{N}-B)$ into the functional integral (\ref{eq8}) at finite 
temperature, where
\beq
\hat{N}=\int\rmd^{2}{\bf{x}}\,{\psi}^{\dagger}({\bf{x}})\psi({\bf{x}})
\,, \hspace{2cm}{\bf{x}}=(x_{1},x_{2})\,\label{eq14}
\eeq
is the fermion number operator. Using an auxiliary scalar field
$\theta(x_{3})$, the above delta-function constraint is exponentiated
and after integrating out the fermions we obtain the partition function
\ba
Z_{f} &=&\int ({\cal {D}}\lambda)({\cal{D}}\theta)\,\exp
\left[ -N\,{\cal{I}}_{eff}(\lambda ,{\cal{G}};\theta,\tilde{B})\right]\,,
\label{eq15}\\ 
{\cal{I}}_{eff}(\lambda,{\cal{G}};\theta,{\tilde{B}})&=&{\rm
i}\tilde{B}\int_{L}\theta(x_{3})\, \rmd x_{3}+ \frac{1}{2{\cal{G}}}
\int_{L} {\rm 
d}^{3}x\,\lambda^{2}   
(x)-\mbox{Tr}[\ln (\slash\!\!\!\partial+{\rm i}\gamma_{3}\theta+\lambda 
)]_{L}  \label{eq16}\,,   
\ea
where $\tilde{B}=B/N$ is assumed to be finite for large-$N$ and
$1/{\cal{G}}$ is the new coupling. The
subscript $L$ denotes $x_{3}$-integration up to $L$ and the latter
quantity  plays here
the r\^ole of inverse temperature $1/T$. 

For large-$N$, the functional integral (\ref{eq15}) can be calculated
by the steepest descent method, since it is dominated by the uniform
stationary points $\llr$ and 
$\lthetar$ of ${\cal{I}}_{eff}$. These stationary points are obtained
as the solutions  of the following set of saddle-point 
equations
\ba  
\frac{1}{{\cal{G}}}
 &=& \frac{2}{L}\,\!\!\sum\limits_{n=-\infty }^{\infty }\int \frac{
{\rm d}^{2}{\bf{p}}}{(2\pi
)^{2}}\frac{1}{{\bf{p}}^{2}+(\omega_{n}+\lthetar)^{2}+\llr^{2}}\,,
\label{first-sp} \\
i\tilde{b} & = & \lim_{\tau \rightarrow 0}
\frac{2}{L} \int \sum_{n=-\infty}^{\infty}\frac{\rmd^{2}{\bf{p}}}{(2\pi)^{2}}
\frac{e^{{\rm i}\omega_{n}\tau}(\omega_{n}+\lthetar)}{
{\bf{p}}^{2}+(\omega_{n}+\lthetar)^{2} +\llr^{2}}\,,\label{second-sp}
\ea 
where $\tilde{b}=(\tilde{B}L/\Omega)$, with $\Omega$ the total volume. The  
regulating term $e^{i\omega_{n}\tau}$ on the r.h.s of (\ref{second-sp}) has 
been discussed in \cite{Marcello} and ensures a finite result in the limit 
$\tau\rightarrow 0$, {\sl{after}} the Matsubara sum has been performed. 
For $\lthetar$ purely imaginary, which corresponds to having a real
chemical potential $\mu=-{\rm i}\lthetar$ \cite{Weiss,Marcello},
(\ref{first-sp}) coincides with a similar saddle-point equation 
obtained in \cite{Kogut,Rosenstein2}. We can   
renormalise (\ref{first-sp}) by substituting for $1/{\cal{G}}$ the bulk 
renormalised coupling $1/G$ from (\ref{eq11}), since the presence of 
$\lthetar$ does not alter its UV behavior. In this way, we can study the 
finite-temperature phase transition of the model in terms of the renormalised 
mass of the bulk fermionic fields and the chemical potential. For example,
(\ref{second-sp}) would now give the critical fermion number
$\tilde{B}_{cr}$ at which space parity is restored. 

However, we are interested here in possible real values of $\lthetar$ which 
satisfy (\ref{first-sp}) and (\ref{second-sp}). The reason is that, if
$\lthetar$ is a real number we can  
set  in (\ref{first-sp}) the coupling $1/{\cal{G}}$ to its bulk critical
value $1/G_{*}$ and obtain  
the following equation for $\llr$
\beq
L\,\llr+\ln\left(1+e^{-L\llr-{\rm i}L\lthetar}\right)
+\ln\left(1+e^{-L\llr+{\rm i}L\lthetar}\right) =0\,.\label{eq18}
\eeq
This has a real solution for $\llr$ in terms of $L$, whenever we have 
$-1\leq\cos\left(L\lthetar\right)\leq -1/2$, or simply 
$2\pi/3\leq L\lthetar\leq \pi$. Note now that $\llr$, which is the
renormalised inverse correlation length, is non-zero for
$L\lthetar\neq 2\pi/3$ corresponding to a finite-size scaling
regime for our fermionic model. Such a regime is absent for the usual
Gross-Neveu model studied in Section 3 and it is a consequence of
keeping the total fermion number fixed. The dimensionless quantity $L\llr$ is
plotted in  Fig. 1 for the allowed values of $\lthetar$.


\begin{figure}[h]
\centerline{\epsfxsize=8cm \epsffile{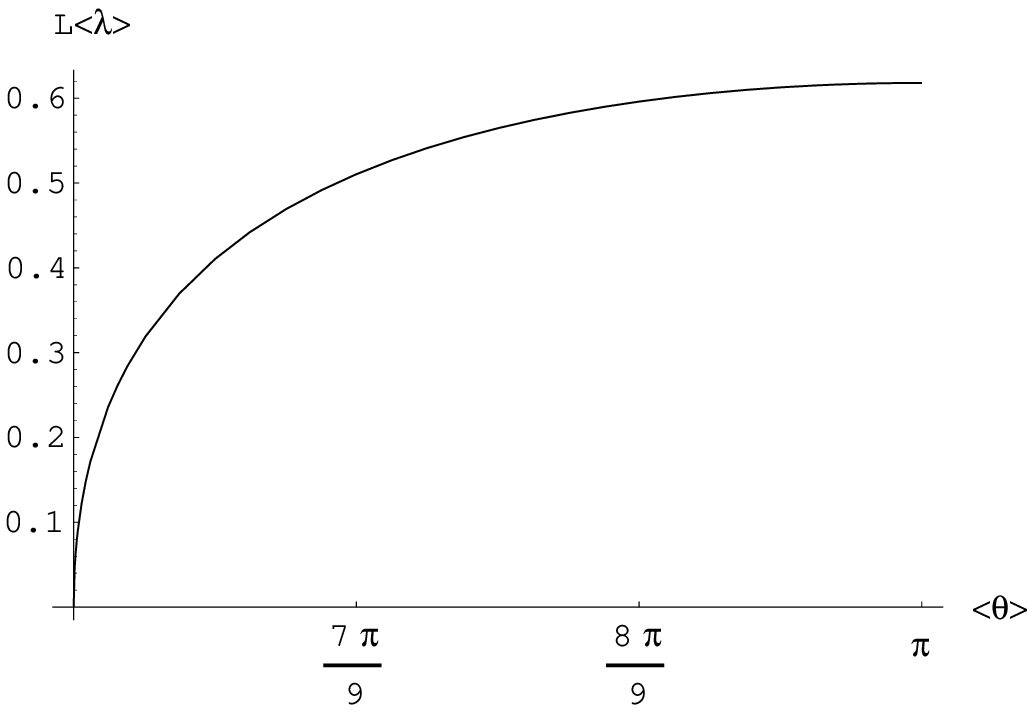}}
\caption{$L\llr$ for the allowed region of $\lthetar$.}
\end{figure}


The second saddle-point equation (\ref{second-sp}) ensures a fixed
mean fermion  
number $\lNr=B$, as imposed by the constraint. It turns out that its r.h.s. is 
real and has a quadratic divergence which should be subtracted. Again, if 
$\lthetar$ is purely imaginary (\ref{second-sp}) assumes the usual form
\cite{Kapusta} of the expression for the conserved charge in a system 
exchanging particles with a reservoir. It also vanishes for 
$\mu \equiv -{\rm i}\lthetar=0$, as it should, since this would correspond to 
the absence of conserved charges. On the other hand, for $2\pi/3\leq
L\lthetar\leq \pi$ we
obtain, after subtraction of the 
divergence, 
\ba
\tilde{b} & =  & \frac{\rm i}{2\pi L^{2}}
\left[Cl_{2}(2\phi)-Cl_{2}(2\phi
-2L\lthetar)-Cl_{2}(2L\lthetar)\right] \,,\label{bcrit} \\
\phi & = & \arctan\left[\frac{e^{-L\llr}
\sin\left(L\lthetar\right)}{1+e^{-L\llr}
\cos\left(L\lthetar\right)}\right]\,,
\ea
where $Cl_{2}(\omega)={\rm Im}\left[Li_{2}(e^{{\rm i}\omega})\right]$ is
Clausen's function \cite{Lewin}. Now, $\tilde{b}$ is related to the total 
fermion number of the system and in principle it should be real and positive. 
Therefore, from (\ref{bcrit}), the only allowed real value for
$\lthetar$ which satisfies both  
saddle-point equations (\ref{first-sp}) and (\ref{second-sp}) is 
$\lthetar =\pi/L$. In this 
case, $\tilde{b}=0$ and it seems that there are no fermions left in the 
system. This is consistent with the apparent bosonization of the theory for 
$\lthetar =\pi/L$ which will be discussed shortly. 
However, it is conceivable that the imaginary solutions for $\tilde{b}$
may also have 
physical meaning, as they give rise to a real value for the  free-energy 
density of the theory. The latter result is rather surprising and is obtained 
by virtue of non-trivial polylogarithmic identities. 

To demonstrate the above points, we calculate the free-energy density 
of the model and we obtain
\ba
\frac{f_{\infty}-f_{L}}{N}
& = & \frac{1}{L}\sum_{n=-\infty}^{\infty} 
\int\frac{\rmd^{2}{\bf{p}}}{(2\pi)^{2}}\ln({\bf{p}}^{2}+
(\omega_{n}+\lthetar)^{2}+\llr^{2})  
-\int\frac{\rmd^{3}p}{(2\pi)^{3}} \ln
p^{2} -\frac{\llr^{2}}{2G_{*}} -{\rm i}\lthetar \tilde{b} \nonumber \\
& = &\, \int\frac{\rmd^{3}p}{(2\pi)^{3}}
\left[\ln\left(\frac{p^{2}+\llr^{2}}{p^{2}}\right)\nonumber 
-\frac{\llr^{2}}{p^{2}} \right]\\
& & 
+\frac{1}{L}\int\frac{\rmd^{2}{\bf{p}}}{(2\pi)^{2}}
\left\{\ln\left(1+e^{-L\sqrt{{\bf{p}}^{2}+\llr^{2}} -{\rm i}L\lthetar}\right)
+\ln\left(1+e^{-L\sqrt{{\bf{p}}^{2}+\llr^{2}} +{\rm i}L\lthetar}\right)
\right\} \nonumber \\ 
& & 
+i\lthetar\int\frac{\rmd^{2}{\bf p}}{(2\pi)^{2}}
\left\{\frac{1}{1+e^{L{\sqrt{{\bf p}^{2}+\llr^{2}}}-iL\lthetar}}-
       \frac{1}{1+e^{L{\sqrt{{\bf p}^{2}+\llr^{2}}}+iL\lthetar}}
\right\} \nonumber \\
& = &
\,-\frac{\llr^{3}}{6\pi} 
+\frac{1}{\pi\,L^{3}}\left[
\ln\left(e^{-L\llr}\right)Li_{2}\left(-e^{-L\llr},L\lthetar\right)-Li_{3}
\left(-e^{-L\llr},L\lthetar\right)\right] \nonumber \\
& &+\frac{L\lthetar}{2\pi L^{3}} \left[Cl_{2}(2\phi)-Cl_{2}(2\phi
-2L\lthetar)-Cl_{2}(2L\lthetar)\right] \,.\label{eq19}
\ea
$Li_{n}(r,\theta)$ is the real part of the polylogarithm
$Li_{n}\left(re^{{\rm i}\theta}\right)$ in Lewin's notation
\cite{Lewin}. As already mentioned, the free-energy density
(\ref{eq19}) is real for $2\pi/3 \leq L\lthetar\leq\pi$.

For $\lthetar=\pi/L$, (\ref{eq18})  
has the solution (\ref{eq5}) $\llr=M_{L}$
(the "golden mean") which  
corresponds to the physical situation of finite-size scaling of the
correlation  
length in the $O(N)$ vector model in $d=3$. This apparent
``bosonization'' of the Gross-Neveu model is effectively a
transmutation between Fermi and Bose  
statistics, since the solution $\lthetar = \pi/L$ introduces  a zero
mode for the fermions. Also, for this particular value 
of $\lthetar$, $\tilde{b}$ is zero which can be
further interpreted
as the absence  of conserved charges in the bosonized version of the 
system. The bosonization of the 
model for $\lthetar=\pi/L$ is also discussed in \cite{Marcello,Farina1} and a 
possible connection with anyonic physics is made in \cite{Farina2}. In this 
case, using the same polylogarithmic identities  as in (\ref{eq6})
we obtain   
\beq
\frac{f_{\infty}-f_{L}}{N}=-\frac{8}{5}\frac{\zeta(3)}{2\pi\,L^{3}}\,,
\label{eq20}
\eeq
which is consistent with the expected CFT result (\ref{eq7}), with
$\tilde{c}/N=-8/5$.  In fact, the full expression (\ref{eq19}) is
consistent with the  scaling form 
(\ref{eq7}) for the allowed values of $\lthetar$, however 
the corresponding expressions for $\tilde{c}/N$ involve polylogarithms
and Clausen's functions at non-exceptional arguments and are not
illuminating. In Fig. 2 we plot the numerical values of $\tilde{c}/N$
corresponding to (\ref{eq19}) for $2\pi/3\leq L\lthetar \leq\pi$.


\begin{figure}[h]
\centerline{\epsfxsize=8cm \epsffile{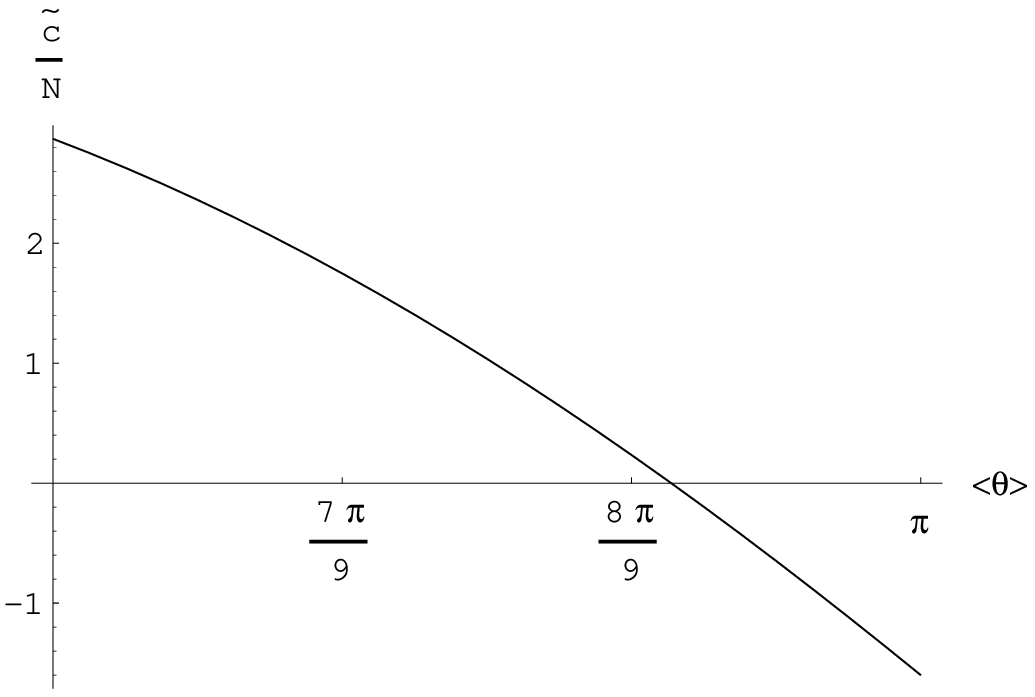}}
\caption{$\tilde{c}/N$ for the allowed region of $\lthetar$.}
\end{figure}


Nevertheless, at the other end-point of the allowed $\lthetar$
region, $\lthetar=2\pi/3L$, (\ref{eq18}) has the solution
$\llr=0$ which corresponds to infinite correlation length. Therefore,
we may associate this point with a temperature phase transition. Remarkably, 
(\ref{eq19}) simplifies considerably for $\lthetar=2\pi/3L$ giving
\beq
\frac{f_{\infty}-f_{L}}{N}=\frac{1}{2\pi L^{3}}
\left[\frac{4\pi}{3}Cl_{2}\left(\frac{\pi}{3}\right)-
\frac{2}{3}\zeta(3)\right]  
\,. 
\label{eq21}
\eeq
It may be interesting to point out that $Cl_{2}(\pi/3)$ is the absolute
maximum of Clausen's function \cite{Lewin} which is a well-documented
numerical constant.  

Consistency of our main result (\ref{eq19}) with the expected from
conformal invariance scaling form
(\ref{eq7}), raises the possibility
that, at the above finite-size critical point with  
$2\pi/3\leq L\lthetar\leq \pi$ the Gross-Neveu model with fixed total  
fermion number is related to conformal field theory. 
This is most clearly seen for $\lthetar =\pi/L$, when from
(\ref{eq20}) one concludes that the model is related to the $O(N)$
vector model at its finite-size scaling critical point. Note 
that the free-energy density given by (\ref{eq20}) is negative which
shows  that such  a critical point is 
unstable by itself.

The interpretation of our results (\ref{eq20}) and (\ref{eq21}) in
terms of the loss of degrees of freedom from the UV to IR is not
clear, except perhaps at the point $\lthetar=\pi/L$ which is related
to the critical point of the $O(N)$ vector model in $d=3$. In
particular, the numerical value of $\tilde{c}$ in (\ref{eq21}) is
$\sim 2.9$ and it is larger than
the sum of $\tilde{c}$'s at the UV fixed 
points of the $\phi^{4}$ theory and the Gross-Neveu-Yukawa theory, when
$N>2.5$. The 
latter sum may be considered as an upper bound for $\tilde{c}$ in 
view of the known universality classes in three-dimensions. Therefore,
the point $\lthetar =2\pi/3L$ corresponds to a genuinely new phase
transition for which the inequalities suggested in \cite{Appelquist}
do not seem to hold. Such a critical point may be associated with a
non-unitary CFT and further with a Lee-Yang edge singularity \cite{TheEdge}. In fact, an interpretation of ${\rm i}\lthetar$ as a
purely imaginary chemical potential suggests  
\cite{newpaper} that, for
$2\pi/3\leq L\lthetar<\pi$, we are dealing with critical points
corresponding to Lee-Yang zeros \cite{Lee-Yang}.

\section{Summary and Outlook}

In this work we calculated the renormalised free-energy
densities of the three-dimensional $O(N)$ vector model and the $U(N)$
Gross-Neveu model, at their conformal points in a geometry with
one finite dimension of length $L$. Our calculations were based on
non-trivial polylogarithmic identities. Our results are consistent 
with the recent suggestion of \cite{Appelquist} that the free-energy density
encodes non-perturbative information regarding the loss of degrees of
freedom from the UV to the IR in QFTs. 

Our study demonstrates the importance of three-dimensional critical
models as testing grounds for the recent ideas concerning
irreversibility of the RG flow and non-perturbative phenomena in QFTs
\cite{Appelquist,Gaite2}. There are many directions in which our
leading-$N$ calculations could be extended. For example, one could
study  the IR limits of the $O(N)$ vector model and 
the Gross-Neveu model in three-dimensions, both in the bulk and in a
finite geometry. These limits would correspond to the lower bounds of 
the inequalities suggested for $\tilde{c}$. The latter could be further
checked by next-to-leading order calculations of $\tilde{c}$ in the
above models\footnote{Similar
studies for the normalisation $C_{T}$ of
the two-point function of the energy-momentum tensor  in the $O(N)$
vector model, have been previously appeared in 
\cite{Tassos*}}. Based on techniques developed here, one could also
study the free-energy density of other three-dimensional models
exhibiting critical behavior, such as $CP^{N-1}$ models \cite{Azakov}
or QED$_{3}$ \cite{Appelquist2}. 

In an possible application of our results, we may view the critical points for
$2\pi/3\leq L\lthetar\leq\pi$ as viable critical points in
supersymmetric CFTs in $d=3$. For example, consider the  
${\cal{N}}=1$ supersymmetric $\sigma$-model in $d=3$ \cite{Gracey2}
which contains Majorana fermions. This means that the large-$N$ fermionic 
contribution to the free-energy density of such a model 
with fixed total fermion number 
is half the r.h.s. of (\ref{eq19}), therefore the sum of the bosonic and
fermionic contributions \cite{Boyanovsky}, as seen from (\ref{eq6})
and Fig.2,  is 
always greater than zero.  This sum vanishes for
$\lthetar=\pi/L$ corresponding to a supersymmetry
restoration mechanism which deserves further study. 

The relevance of non-trivial polylogarithmic identities  to the calculation of 
free-energy densities  in two-dimensional CFTs is well-known
\cite{Jose}. Their appearance in studies of CFTs  
in higher dimensions is intriguing and  
requires further investigation. It would also be interesting to study the OPE 
structure of correlation functions in the Gross-Neveu model for fixed
total fermion number, at the above critical points. 
Although our results simplify for $d=3$, they can presumably 
be generalised for all $2<d<4$. 

\small
\subsection*{Acknowledgments}
A. C. P. would like to acknowledge the hospitality of the ``Erwin
Schr\"odinger Institute'' in Vienna,  where this work started. He would 
also like to thank Prof. I. Todorov for discussions and H. Osborn for
a critical reading of the manuscript.
This work was partially supported by CAPES, a Brazilian agency for the 
development of science and by DAAD A/98/19026 Fellowship.

\end{document}